# Nanoscale Self-Healing Mechanisms in Shape Memory Ceramics

Ning Zhang and Mohsen Asle Zaeem[*]

Department of Mechanical Engineering, Colorado School of Mines, Golden, CO 80401, USA

**Abstract**

Shape memory (SM) ceramics, such as yttria-stabilized tetragonal zirconia (YSTZ), are a unique family of SM materials that offer unique properties including ultra-high operating temperature, and high resistance to chemical corrosion and oxidation. However, formation of defects is usually observed in SM ceramics during manufacturing and/or by mechanical deformation. To fully take advantage of the SM properties of these ceramics, it is necessary to fully understand the nano-structural evolution of defects under external stimuli. In this study, defect closure behaviors in YSTZ nanopillars are investigated by atomistic simulations. Two characteristic orientations of $[01\bar{1}]$ and $[001]$ are selected to represent the dominant deformation mechanisms of phase transformation and dislocation migration, respectively. With the presence of crack and void, the strength and yield strain of nanopillars are noted to decrease significantly, especially for $[01\bar{1}]$-oriented YSTZ nanopillars. Volume expansion associated with the tetragonal to monoclinic phase transformation is observed to promote healing of crack and void. Atom stress analyses reveal stress concentrations along the newly formed monoclinic phase bands. A critical crack width is identified, less than which the crack can be fully closed in compression. Size effect study reveals that an increase in nanopillar size has a positive effect on crack self-healing behavior. For [001]-oriented YSTZ nanopillars, dislocation migration leads to formations of an amorphous phase, which also assist the crack and void closure process. The revealed crack/void healing mechanisms may provide a path for mitigating internal defects that influences the mechanical properties and deformation mechanisms of SM ceramics.

[*] Corresponding author. Email address: zaeem@mines.edu

**Introduction**

Shape memory (SM) ceramics are a unique family of SM materials with a wide variety of applications, such as ultra-high energy dissipation and high-temperature actuation. SM ceramics offer many advantages compared with SM metallic alloys, such as higher strength, higher operating temperature, better thermal stability, and superior oxidation/corrosion resistance. Among SM ceramics, zirconia ($ZrO_2$) based ceramics have been the focus of substantial investigations in the past few decades [1-3]. While it is of growing interest for a variety of electrical, thermal, chemical and other related applications, one of the most common motivation for zirconia development is to utilize its great mechanical performance. The superior properties of zirconia based ceramics, such as high strength and fracture toughness, good wear resistance, and low coefficient of friction, have made them suitable for stress-bearing structural and implant applications [4,5]. In particular, the recently revealed remarkable SM property of zirconia based nanopillars [6,7] has triggered new research interests of SM ceramics.

Zirconia exists in three crystalline phases. The monoclinic phase is stable at room temperature, the tetragonal phase exists between ~1,400 K and ~2,500 K, and the cubic phase is the high-temperature phase which exists above ~2,500 K [3]. By cooling the pure tetragonal zirconia down to a temperature lower than 1,400 K, the first-order, displacive tetragonal to monoclinic ($t{\rightarrow}m$) phase transformation is activated. This type of transformation is associated with shape deformation and a volume expansion of about 4.0%, which will lead to undesired residual stresses and consequently results in crack and porosity formation, with detrimental effects on structure integrity and durability of the zirconia substrate [8]. By doping various oxides, e.g. yttrium oxide or yttria ($Y_2O_3$), the tetragonal phase can be retained in a stable state at room temperature [3]. Recently, a crystal orientation dependence of martensitic transformation in yttria and titania stabilized zirconia ceramics was revealed in an experimental study [9]. Meanwhile, by using large scale molecular dynamics (MD) technique, the plastic deformation mechanisms of yttria-stabilized tetragonal zirconia (YSTZ) was identified to have a strong dependence on the orientation of the crystal structures [10]. Dislocation migration and phase transformation behaviors were observed when the YSTZ nanopillars were compressed along different orientations. And the high strength of YSTZ ceramics was found to be derived from the stress-induced $t \rightarrow m$ phase transformation [9,10]. The nanoscale mechanical response of YSTZ nanopillar is also shown to be sensitive to its specimen size and yttria concentration [11]. The formation of oxygen vacancies along with the addition of yttria was found to play a vital role in influencing the strength of nanopillars. Additional studies on interface deformation of bicrystal YSTZ [12] indicated that $t \rightarrow m$ phase transformation preferably initiates from the grain boundary (GB), and also the highest strength of nanopillar is obtained when loading direction is parallel to GB. Volume expansion associated with martensitic phase transformation is responsible for the observed strengthening behavior.

In the manufacturing industry, fine polishing and grinding are used as the most productive techniques for finishing tetragonal zirconia components [13,14]. With the dramatic improvement of modern manufacturing techniques, although the bulk defects have been significantly reduced, ground workpieces are often left with cracks, pores and a limited amount of plastic deformation due to the hard and brittle nature of zirconia ceramics. Such subsurface damage and residual stresses may seriously alter the material properties, and cause strength degradation or even a catastrophic failure of the ceramic components. Besides, formation of defects was also commonly observed in deformed zirconia ceramics [15]. An interesting question arises that how these defects evolve under external loading? It is well known that the $t \rightarrow m$ phase transformation can be locally triggered, either mechanically by the presence of high stress [16,17] or chemically by the diffusion of water species from the environment [5]. Experimental studies have revealed that under mechanical loading the accompanied volume expansion with phase transformation may close the crack tip, and thus prevent further crack propagation to achieve high strength and fracture toughness, i.e., phase transformation induced toughening [3]. In most defected brittle materials, fracture is usually initiated due to the stress concentration at flaws, such as cracks [18,19] and voids [20]. Such stress concentration regions in zirconia based ceramics may act as a potential source of phase transformation and dislocation nucleation.

The concept of crack closure was first introduced by Elber [21,22] to explain the characteristics of fatigue crack growth in aluminum alloys. Originally, closure displacements arising from the crack tip plasticity were considered to be the major factors. Since then, a number of models based on crack closure have been postulated. For example, an elastic phase-field model was proposed recently to capture the effect of external stress on the $t \rightarrow m$ phase transformation in zirconia [23-25]; under uniaxial tensile loading, the process of a static elliptic crack closure during the $t \rightarrow m$ phase transformation was elucidated [26]. However, the fracture behavior that involves dislocation emission from cracks is still an open problem of both fundamental and technological relevance; this is a complex process of nucleation of a dislocation from crack surfaces and its subsequent movement away from the crack [27].

Thera are a few nanoscale studies to investigate the crack propagation and healing in metals. For example, a new mechanism that leads to complete healing of nanocracks in nanocrystalline metals was discovered by MD simulations [28]; Disclinations by migrating GBs was noted to be responsible for crack closure. Furthermore, the crack healing mechanism of a copper nanopillar under shearing loading was also investigated recently; dislocation shielding and the atomic diffusion was revealed to control the crack closure [6]. It is important to note that vacancies, as important stress concentration originators, also interact with dislocations to alter the stress field around defects [29,30]. Despite the recognized importance of the stress field in the vicinity of a crack or a vacancy on dislocation emission, its behavior in SM ceramics, in particular in zirconia based ceramics, has not been studied directly at the atomic scale.

Although some progress has been made in terms of understanding the underlying mechanisms of transformation toughening of zirconia ceramics and crack/void closure behavior in metals, a comprehensive atomic-scale understanding of the influences of *t→m* phase transformation on evolution of nano-cracks and nano-voids in YSTZ ceramics is still lacking. The evolution and subsequent interaction of nano/microstructure with a nano-crack/void is amongst the most basic problem in the field of materials science and engineering. The nanostructural features, such as phase transformation patterns and dislocations, have pronounced effects on the crack/void propagation behavior by altering the stress field around the crack/void.

To this end, this paper is aiming at investigating the effects of phase transformation and dislocation migration on crack/void healing in 4.0 mol% YSTZ nanopillars under compressive loading. A visible crack/void closure process will be presented along with the detailed discussions of governing mechanisms. Crack/void closure generally involves two stages: the mechanical closure of crack/void, i.e., reducing the volume, and the final bonding of internal (crack/void) surfaces providing complete healing and thus a sound material. The present work will focus on the first stage only: the mechanical closure.

## Results and Discussion

*Effect of phase transformation on crack/void closure*

Under uniaxial compression the opening of a pre-existing crack or void occurs usually perpendicular to the loading direction if only elastic deformation is involved. In this section, the influence of phase transformation during plastic deformation on structural evolution of cracks and voids will be systematically studied. Our recent study [10] showed that $[01\bar{1}]$ is one of the orientations along which phase transformation is the dominant deformation mechanism; accordingly, $[01\bar{1}]$-oriented YSTZ nanopillar models with pre-existing crack and void are built (Fig. S1) to study the phase transformation effect on internal crack/void closure behavior during uniaxial compression along $[01\bar{1}]$ direction.

To investigate the crack size effect on mechanical properties and crack closure behavior of YSTZ nanopillars, four typical cracks with different length (*2a*) and width (*2c*), but the same length/width ratio (*a/c* = 7.0) are created, i.e., 2*c* = 0.8 nm, 1.4 nm, 2.0 nm, and 2.6 nm. The obtained stress-strain (*σ* - *ε*) relations of $[01\bar{1}]$-oriented YSTZ nanopillars with different crack sizes under compression are presented in Fig. 1a. For comparison purposes, the mechanical response of defect-free YSTZ nanopillar is also shown. It can be seen from Fig. 1a that with the presence of cracks, the strength and yield strain decrease about 9.0% and 50.0%, respectively, compared with those of the defect-free YSTZ nanopillar. The nanopillar with crack width of 2.6 nm shows the lowest strength and yield strain. For all the nanopillars with pre-existing cracks, after the major sudden stress drop at *ε* ≈ 0.6%, an increase of plastic stress is observed in the strain region of 0.6% < *ε* < 1.2%, which is not observed for the defect-free YSTZ nanopillar.

Nanovoids with diameters ($d$) of 2.0 nm, 4.0 nm, 6.0 nm and 8.0 nm are introduced into the [01$\bar{1}$]-oriented YSTZ nanopillars to study the void size effect on their mechanical responses and void closure process. The stress-strain relations are plotted and compared with the defect-free nanopillar in Fig. 1b. It is interesting to note that with the presence of internal voids, the overall trend of stress-strain curves is dramatically changed, with a significant decrease of strength (~60%) and Young's modulus compared to the defect-free nanopillars. These characters are different from the studied cases of YSTZ nanopillars with pre-existing cracks, which suggests that the presence of void may result in a different phase transformation behavior. Regarding the void size effect, it is noted that as the diameter of void increases from 2.0 nm to 8.0 nm the yield strain of YSTZ nanopillar decreases from 0.6% to 0.3%. The stress corresponding to the region of $\varepsilon > 1.0\%$ decreases with the increase of void diameter.

Figure 2 illustrates the effect of phase transformation on crack closure in nanopillars with the aforementioned four crack sizes in Fig. 1a. It is revealed that two phase transformation bands form around the crack. With continuous loading, the transformed monoclinic phase bands (green atoms in Fig. 2) becomes thicker. Meanwhile, the experimental explored phenomenon of volume expansion associate with phase transformation [3] is also observed, which leads to the trend of closure of the internal cracks. However, it is worth to mention that the volume expansion by $t \rightarrow m$ phase transformation was observed for $\varepsilon < 4.0\%$ according to experimental measurement [3], therefore once the tetragonal phase around a crack transforms entirely to the monoclinic phase, the specimen will lose the ability of sequential crack closure.

The crack width is tracked and recorded during compression. Since the crack width varies in different nanopillars, we define a normalized crack width as: $2c'/2c$, where $2c'$ and $2c$ are the instantaneous and original width of the crack, respectively. $2c'/2c = 1.0$, $2c'/2c > 1.0$ and $2c'/2c <$ 1.0 stand for the status of invariability, opening and closure of the pre-existing cracks, respectively. A crack is considered to be at a stable condition when $2c'/2c \leq 1.0$ during compression. A fully closed crack is obtained when $2c'/2c$ decrease to zero. In Fig. 3a, the normalized crack width versus applied strain ($2c'/2c$ versus $\varepsilon$) is plotted. For an internal crack under compression along its long axis, it is expected that the crack opens perpendicular to the loading direction when merely elastic deformation is involved. For all the studied cases, the cracks open ($2c'/2c > 1.0$) in the elastic region, which is denoted as "elastic dilation of crack" region in Fig. 3a. Subsequently, the crack width decreases rapidly ($2c'/2c < 1.0$). For the nanopillar with $2c = 0.8$ nm, crack closes very fast, and the normalized crack width ($2c'/2c$) drops to around 0.05 when the strain increases to 2.8%. The value of $2c'/2c$ fluctuates around a constant value of 0.05 (Fig. 3a), indicating 95% crack closure is achieved, which can be also observed in Fig. 2a''. For nanopillars with $2c = 1.4$ nm or 2.0 nm, the $2c'/2c$ also drops to constant values around 0.42 and 0.6 (see Fig. 3a), respectively, which implies 58.0% and 40.0% crack closure. However, for the case of $2c = 2.0$ nm, the value of $2c'/2c$ slightly increases for $\varepsilon > 6.0\%$, which implies reopening of the crack after phase transformation around the crack is completed. When crack width increases to 2.6 nm, $2c'/2c$ first decreases to ~0.7 at 4.0% strain, then it increases to 1.0 with continuous loading. Such phenomenon suggests that as the crack width increases beyond a critical value, in this study $2c_0 \approx 2.0$ nm, the volume expansion caused by $t \rightarrow m$ phase transformation becomes incapable of closing the pre-existing internal cracks. It is also worth mentioning that for $2c = 2.0$ nm (Fig. 2c-c'') and $2c = 2.6$ nm (Fig. 2d-d''), the original tetragonal phase does not transform to monoclinic phase in the regions around the top and bottom of crack tips. In contrast, the expansion of transformed monoclinic phase band covers all the surrounding area of cracks when $2c = 0.8$ nm (Fig. 2a-a'') and $2c = 1.4$ nm (Fig. 2b-b'').

The effect of specimen size on the nanostructure evolution around the crack is also studied (Fig. S2). To systematically analyze the crack closure process, we summarize and plot the relation of normalized crack width ($2c'/2c$) and strain in Fig. 3b. It is noted that the crack self-healing behavior caused by $t{\rightarrow}m$ phase transformation becomes more effective with the increase of the width ($L$) of nanopillars. For the nanopillars with $L$ = 20.0 nm and 30.0 nm, $2c'/2c$ first drops to minimum values of 0.7 and 0.55, indicating 30% and 45% crack closure (Fig. 3b), respectively. Later, $2c'/2c$ increases back to 1.0 and 0.8, implying crack re-opening. On the other hand, $2c'/2c$ drops to around 0.3 at the strain of 5.8% in nanopillars with $L$ = 40.0 nm and 50.0 nm, then the crack width stays in a relatively steady state condition. Such observation suggests that $t{\rightarrow}m$ phase transformation may play a significant role in healing the internal cracks of bulk YSTZ pillars.

To track the phase transformation behavior around the pre-existing void, we present the atomic trajectories at various strains in Fig. 4. It is revealed that two transformed monoclinic phase bands form around each nanovoid, and then expand along the loading direction. The size of void has a remarkable effect on the path of phase transformation. The new phase is observed to finally expand across the nanovoid when $d$ = 2.0 nm (Fig. 4a''). However, after the full closure of the void in nanopillar with $d$ = 4.0 nm, part of the tetragonal phase around the void refuses to transform to monoclinic phase (Fig. 4b''). Additionally, $t{\rightarrow}m$ phase transformation occurs in the left side and right side regions of void rather than the bottom and top regions. Similar to the aforementioned normalized crack width, to quantitatively analyze the evolution of void closure versus applied strain (Fig. 4e), we define a normalized diameter ($d'/d$), where $d$ is the initial diameter and $d'$ is the instant width of void which is measured perpendicular to the loading direction during compression. It is noted that the region of elastic dilation of void becomes shorter compared to the YSTZ nanopillars with pre-existing cracks (shadow region in the left side of Fig. 4e). In particular, no obvious elastic dilation of void is observed in the nanopillars with $d$ = 6.0 nm and 8.0 nm. For the cases of $d$ = 2.0 nm and 4.0 nm, $d'/d$ reduces rapidly due to the volume expansion by phase transformation, and reaches zero when the compressive strains become 3.2% and 4.5%, respectively, indicating that the voids are fully closed. However, as $d$ increases to 6.0 nm and 8.0 nm the value of $d'/d$ remains around 1.0, implying no reduction in the width of void. Such phenomenon can be explained by the restrained phase transformation region around the void in Fig. 4c''-d''. Considering the severe stress concentration on the left and right side edges of voids, it is reasonable to infer that specimen size has insignificant effect on the void closure process, therefore it is not discussed.

The atom stress along the loading direction ($\sigma_{yy}$) is plotted in Fig. 5. As representative, we show the stress contour of the nanopillar with a crack width of 2.6 nm in Fig. 5b. Comparing to the phase transformation map in Fig. 5a, it can be noted that the distribution of $\sigma_{yy}$ is highly non-uniform, and the negative compressive stress concentrates along the band of monoclinic phase. Such pattern arises from the sudden crystal structure change (from tetragonal to monoclinic). The compressive stress can be explained by the observed volume expansion normal to the crack surface direction. Similar pattern of atom stress distribution is observed in the other YSTZ nanopillars with different crack sizes. We also show the normal atom stress distribution of a nanopillar with a pre-existing void in Fig. 5b. It is clear that the stress concentration around the void is more severe and localized compared with that of the large band region around the crack in nanopillars with pre-existing cracks (see Fig. 5a and Fig. 2). This is because the left and right side curvatures of void are much larger than those of crack. Hence, phase transformation is much easier to initiate from the left and right side edges of void than from those of the crack, which accordingly lead to the dramatic drop of strength in Fig. 1b. Furthermore, the transformed monoclinic phase band contact with the void at a very limited area, as shown in Fig. 4c'', d'', which directly impair the degree of void healing by phase transformation.

*Effect of dislocation migration on crack/void closure*

Besides phase transformation, our recent study [10] also showed that in some specifically oriented nanopillars, such as [001]-oriented YSTZ nanopillar, dislocation migration dominates the plastic deformation of YSTZ nanopillars. In this section, we perform compression simulations of [001]-oriented YSTZ nanopillars with pre-existing cracks and voids to investigate the effect of dislocation migration on crack/void closure.

The length/width ratio of pre-existing cracks is still controlled at 7.0 similar to the previous section. Although the strength and yield strain decrease due to the appearance of vertical central cracks (stress-strain curves in Fig. S3a), the equilibrium stresses after plastic flow, corresponding to $\varepsilon > 1.0\%$, are equivalent or even slightly increased compared to the defect-free specimen. To track the evolution of crack closure due to the occurrence of dislocation glide, we present the side view of atomic configurations of deformed YSTZ nanopillars in Fig. 6a-c. It can be seen that with the propagation of dislocation the width of the crack reduces rapidly. To better understand the dislocation motion behavior around crack, a slip vector code [31] is also used to extract the dislocation glide planes in Fig. 6a'-c'. It is worth to mention that there is an insignificant crack size effect on the crack healing process (Fig. S3a), hence only the trajectories of one of the cases (the nanopillar with crack width of 1.4 nm) are presented in Fig. 6 to represent the dislocation dominant deformation process. For this case study, the YSTZ nanopillar remains elastic until ~ $\varepsilon$ = 0.5%, at which point dislocation starts to nucleate from the crack surface and then propagate across the crack along the (111) primary slip plane (see Fig. 6a'). Subsequently, the stacking fault becomes thicker and thicker due to the successive gliding of partial dislocations, as shown in Fig. 6a'-c'. On the other hand, secondary dislocation nucleation and propagation occur. The stress fluctuations in the plastic flow region (Fig. S3a) are caused by the accumulation of stacking faults. As illustrated in Fig. 6a-c, with the nucleation and migration of dislocation, the width of crack reduces, and crack healing can be observed. This is because both the left- and right-side surfaces of the crack suffer from a large plastic deformation, which is accommodated by evolution of dislocations and dislocation-crack surface interactions. Compared with defect-free systems, pre-existing defects result in an increment in the free energy of the whole system, however this contribution is not sufficient to trigger the defect closure process. Some previous experimental researches [32,33] suggest that the main barriers for crack healing are the crack surface tension and the activation energy for recrystallization around the crack surface. In the current case study, dislocation slip occurs around the middle region of crack due to the local stress concentration. Consequently, the nucleation and migration of dislocation overcome the barriers for crack surface shearing and recrystallization around the crack surface, and serve as the main driving force for crack closure.

Compared to the case of defect-free [001]-oriented YSTZ nanopillar, with the appearance of voids, strength and yield strain are noted to decrease significantly (about 16~35% and ~67% reductions, respectively), while Young's modulus remains at a similar value (Figure S3b). However, comparing to the case of phase transformation effect on void healing (Fig. 1b), the current decrease in strength is much smaller. Also, the size of voids shows more significant effect on the strength of the nanopillar; 19% reduction of strength is observed when the diameter of nanovoid increases from 2.0 nm to 6.0 nm.

It is revealed that two dislocations emit and propagate rapidly around each void due to the high stress concentration (Fig. S4), resulting in the first major sudden drop in stress (Fig. S3b). As the compressive strain increases continuously, more and more partial dislocations are triggered, which lead to the thickening of stacking faults and formation of localized amorphous structure (see Fig. S4d', d''). Besides, dislocation bifurcations are induced from the primary slip plane. Meanwhile, the dislocation nucleation and migration around void results in the decrease of void opening displacement, and the shape of the original void has changed from circular to almost rectangular shape (Fig. S4b'' and Fig. S4c''). The void closure in [001]-oriented YSTZ nanopillar is facilitated by the local dislocation nucleation and migration, as well as amorphous phase formation.

In summary, the magnitude of stress around defects is magnified, thereby triggering phase transformation at a lower applied load and a smaller deformation strain with respect to defect-free YSTZ nanopillar. Among all the studied nanopillars with pre-existing cracks and voids, the most significant decrease of strength occurs in the $[01\bar{1}]$-oriented YSTZ nanopillars with pre-existing voids due to the localized stress concentration around the void edges. In YSTZ nanopillars with phase transformation as the dominated deformation mechanism, the closure of cracks and voids was observed due to the volume extension accompanied by phase transformation. Crack size effect study revealed that a critical crack width exists, below which the crack can be fully closed under compression. When the crack width increases to larger than the critical value only partial crack healing can be achieved due to the limited volume extension (~ 4 %) by phase transformation. Atom stress analyses disclosed that the negative compressive stress concentrates along the new transformed monoclinic band. Specimen size also shows noteworthy effect on crack closure process; the crack self-healing behavior becomes more effective with the increase of nanopillar width, which suggests that phase transformation may play a significant role in healing the internal cracks of bulk YSTZ pillars. The size of void shows remarkable effect on the path of phase transformation; phase transformation prefers to occur perpendicular to the compressive loading direction. For YSTZ nanopillars with dislocation migration as the dominated deformation mechanism, it is noted that the void size has significant effect on the mechanical response and void healing, while the crack size has negligible effect. Dislocation nucleation and migration lead to the formation of stacking faults and amorphous phase, which mediate crack and void closure.

In some appropriate circumstances YSTZ ceramics can be self-healed by mechanical loading to eliminate their internal defects. The findings of this work will provide insight into the possibility of tuning the crack and void sizes to elicit desired mechanical properties of SM ceramics, and also to guide the design of the appropriate structural components with self-healing ability in order to obtain longer service life.

**Methods**

MD simulation is a very powerful computational technique in investigating the nucleation and evolution of plastic deformation and associated failure mechanisms at the atomistic scale. This approach yields very detailed atomic information about the simulated system. And it has been applied extensively to study the fundamental mechanisms of failure behavior of materials in the past, including the problems of phase transformation [10,34,35] and twinning [35,36], dislocation migration [10,34], and crack propagation [37]. In the current work, the interaction between atoms in YSTZ nanopillar is simulated within the framework of short-range Born-Meyer-Buckingham (BMB) potential and the long-range Coulomb potential [38], which has been demonstrated to property model both tetragonal and monoclinic zirconia phases. The functional format and detailed parameters of the employed potential were given in our previous study [10].

The MD simulation geometry of single crystalline YSTZ nanopillars containing a central vertical crack or a void are shown in Fig. S1. Large-scale Atomic/Molecular Massively Parallel Simulator (LAMMPS) program is utilized in this work for MD simulations [39]. The Nose-Hoover thermostat [40] is applied to maintain the system temperature at a constant value of 298 K. Since the plastic deformation mechanisms of YSTZ nanopillars are strongly dependent on their crystallographic orientations [9,10], two characteristic orientations of $[01\bar{1}]$ and $[001]$ along the loading direction are selected to build the models to represent dominant deformation mechanisms of phase transformation and dislocation migration, respectively. The representative YSTZ nanopillar has a dimension of $10 \times 20 \times 47$ nm³ with approximately 800,000 atoms; dimensions vary for the study of specimen size effect. The nano-crack/void is generated by removing the corresponding atoms in the center of the YSTZ nanopillar. Free surface boundary conditions are employed along all three directions. For the compression simulations, a uniform strain is imposed on the top and bottom surfaces with a total strain rate of $\dot{\varepsilon} = 2 \times 10^7 \ s^{-1}$. Since the introduction of crack and void creates new surfaces in the internal of computer model, a full relaxation was performed before the compressive loading is applied. The time step is set to be 1 femtosecond.

*Atomistic stress definition*

Stress calculation at atomic scale has been a subject of theoretical debate, basically because there is a lack of linkage between the atomistic stress formulation and the fundamental concept of Cauchy stress, which is the actual physical quantity measured in experiments [41]. In quantum and classical molecular mechanics, the local stresses are usually calculated based on virial theorem, which is the most commonly used definition of stress in discrete particle systems [42]. As shown in Eq. (1), virial stress contains two parts, kinetic energy part due to the motion of individual atoms and potential energy part due to the interaction between atoms. Since the virial stress in Eq. (1) is written as a sum over atoms, the local atomistic stress can be considered as the individual term in the formula, which is usually referred to *atomic virial stress* [41]. This is the stress formula that has been exclusively employed in MD simulations, e.g., LAMMPS uses the virial theorem to calculate atomistic stress tensors [43]. Another expression for the atomistic stress is based on the concept of Cauchy stress. Different from the volume-averaged virial stress, this stress formulation is constituted by simply area-averaged forces, as can be seen in Eq. (2) [44]. Although it was claimed that this area-averaging formulation can remove "unphysical" stress oscillations near the ends of a free surface, it leads to a loss of symmetry of the stress tensor. Apparently, the stress calculated through the area-averaged formula is identical to the experimental measured stress when thermal stress is negligible [45]. In other words, this formulism is applicable when thermal stress is ignored. However, in the current study the operating temperature is controlled at 298 K, which makes significant contribution to the calculated atomistic stress. And also, in order to produce symmetric stress tensors, virial stress formalism is adopted.

$$\sigma_{ij} = -\frac{1}{V}\sum_{\alpha}\left(m^{\alpha}v_i^{\alpha}v_j^{\alpha} + \frac{1}{2}\sum_{\beta \neq \alpha}^{N} F_i^{\alpha\beta}r_j^{\alpha\beta}\right) \qquad (1)$$

$$\lambda_{ij} = \frac{1}{A}\left[\frac{1}{2}\sum_{\alpha}\frac{m^{\alpha}v_i^{\alpha}}{\Delta t} - \frac{1}{2}\sum_{\alpha}\sum_{\beta}\frac{\partial V}{\partial r^{\alpha\beta}}\frac{r_i^{\alpha\beta}r_j^{\alpha\beta}}{|\vec{r}_j^{\alpha\beta}|}\right] \qquad (2)$$

where $i, j = 1, 2, 3$ indicate the $x$, $y$, and $z$ directions; $m^{\alpha}$ and $v^{\alpha}$ are the mass and velocity of atom $\alpha$; $F^{\alpha\beta}$ is the force on atom $\alpha$ resulting from the pair interaction with atom $\beta$. $r^{\alpha\beta}$ is the distance between atoms $\alpha$ and $\beta$.

It is worth emphasizing that the volume of individual atoms is not well defined or not easy to compute in a deformed solid or liquid. Although LAMMPS proposes a possible way to estimate the per-atom volume through calculating the Voronoi tessellation of the atom, this method may lead to underestimation of Voronoi volumes in low density systems. Besides, it is very computationally expensive and may output unnecessary information of per-atom volume at some insignificant deformation sequences. In general, the atom stress is calculated through post processing, and per-atom volume is taken as the average value of the total simulation box volume, i.e.,

$$V = \frac{V_{total}}{N} \qquad (3)$$

where $V_{total}$ and $N$ are the volume of the simulation box and total number of atoms, respectively. Such estimation hypothesizes that the simulation model maintains as a homogeneous system, and it works well under the situation that no significant changes of local atomic density occur during the deformation process. For example, the average volume per atom in a stress-free Pd crystal model is used to track the atom stress distribution during crack propagation [46]. However, such estimation of atom volume is incapable of accurately predicting the atom stress distribution when there are significant changes of local atomic density, such as in phase transformation, partial dislocation migration and amorphous phase formation. In Fig. S5, we take tetragonal to monoclinic phase transformation in YSTZ as an example to illustrate the change in local atomic density in different phases and regions, which apparently demonstrates the limitation of average per-atom volume. It is clear that the phase transformation has a noteworthy influence on the local atom density and accordingly on per-atom volume, especially around phase boundary and in the regions with new phase and amorphous phase.

We propose a new method to calculate the per-atom volume by introducing a cut-off distance ($R$) around the target atom $\alpha$. The number of atoms in this spherical region is counted and denoted as $N$. Then the atom volume of atom $\alpha$ can be obtained by:

$$V_\alpha = \frac{\Omega}{N} = \frac{4}{3N}\pi R^3 \tag{4}$$

where $\Omega$ is the volume of the spherical cut-off region. It should be noted that the cut-off distance should be larger than the nearest neighbor distance but smaller than the third nearest neighbor distance. In the current study, $R = 3.5\,\text{Å}$ is chosen. The advantages of our proposed method are twofold; it can estimate the per-atom volume more accurately, and also the post processing target makes it computationally efficient. Furthermore, it is required that the sum of atom volume should satisfy the total volume conservation condition, $\sum V_i = V_{total}$ [47]. Therefore, the calculated atomistic stress $\sigma_{ij}$ will be rescaled by multiply a ratio of $V_{total}\Big/\sum_{\alpha=1}^{N} V_\alpha$ to obtain the real atomic virial stress.

**Acknowledgements**

This work was supported by the U.S. Department of Energy, Office of Science, Basic Energy Sciences, under Award number DE-SC0019279. The authors are grateful for computer time allocation provided by the Extreme Science and Engineering Discovery Environment (XSEDE).

**Data Availability**

The data that support the findings of this study are available from the corresponding author upon reasonable request.

**Author Contributions**

N.Z. and M.A.Z. wrote the manuscript. N.Z. carried out the MD simulations, and M.A.Z. coordinated the whole work.

**Competing Interests:** The authors declare no competing interests.

# Figures

# Nanoscale Self-Healing Mechanisms in Shape Memory Ceramics

Ning Zhang and Mohsen Asle Zaeem[*]

Department of Mechanical Engineering, Colorado School of Mines, Golden, CO 80401, USA

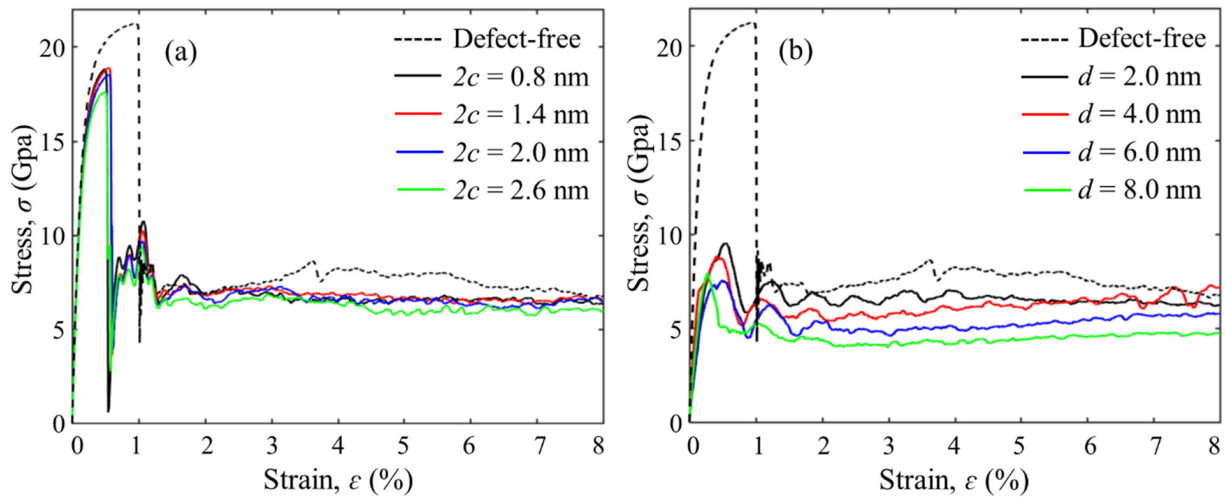

Figure 1. Stress-strain ($\sigma$ - $\varepsilon$) relations of $[01\bar{1}]$-oriented YSTZ nanopillars (a) with and without cracks, and (b) with and without voids. The length/width ratio of cracks is controlled at 7.0. The crack widths are 0.8 nm, 1.4 nm, 2.0 nm and 2.6 nm. The voids have diameters of 2.0 nm, 4.0 nm, 6.0 nm and 8.0 nm.

---

[*] Corresponding author. Email address: zaeem@mines.edu

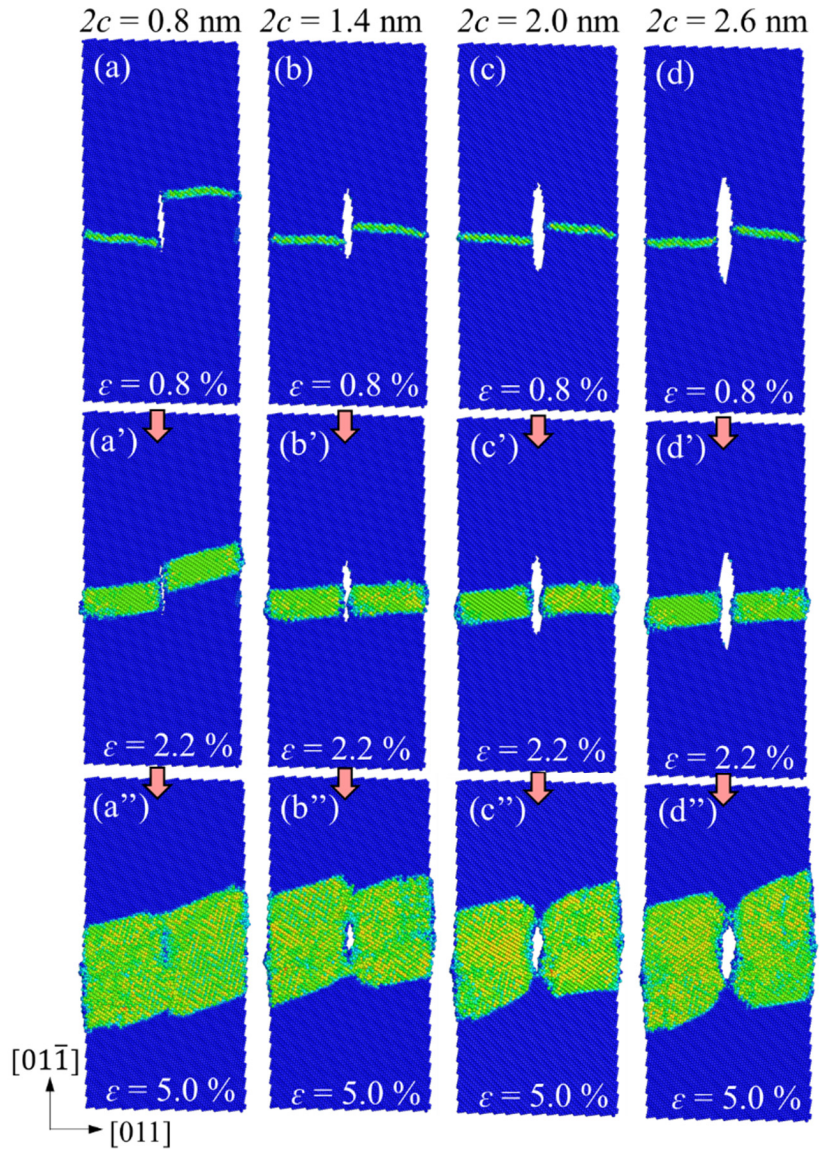

Figure 2. [01$\bar{1}$]-oriented YSTZ nanopillars with pre-existing crack under compressive strains of (a-d) 0.8%, (a'-d') 2.0%, and (a''-d'') 5.0%. The length/width ratio ($a/c$) of crack is 7.0. (a-a''), (b-b''), (c-c''), and (d-d'') illustrate the plastic deformation process of YSTZ nanopillars with crack width of $2c$ = 0.8 nm, 1.4 nm, 2.0 nm, and 2.6 nm, respectively. The atoms are colored by coordination number, where blue and green colors denote atoms with tetragonal and monoclinic crystal structures, respectively.

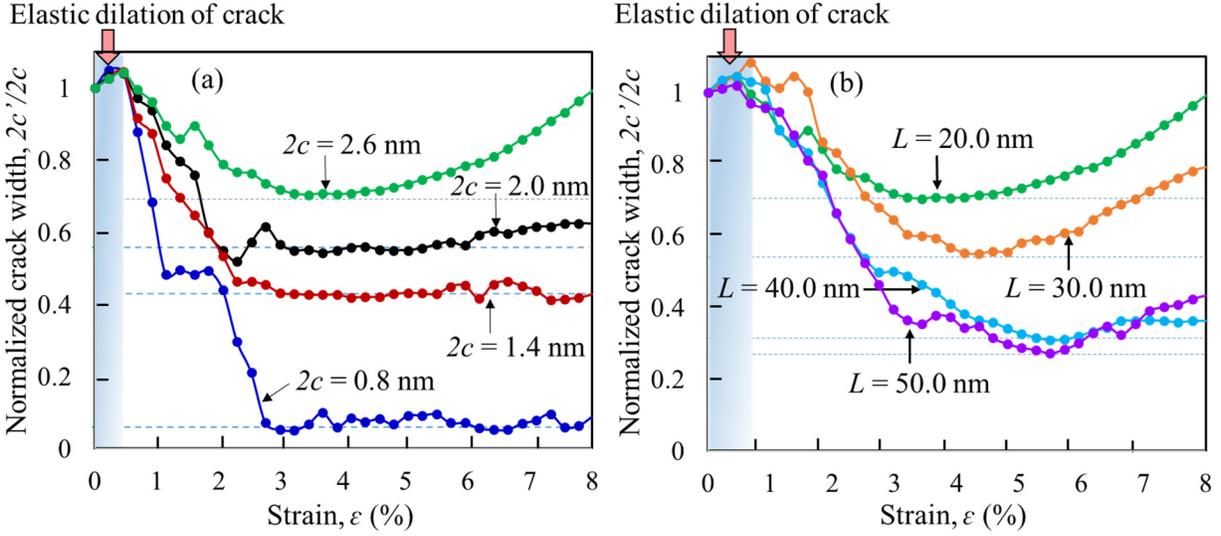

Figure 3. Evolution of normalized crack width ($2c'/2c$) during compression as a function of strain for YSTZ nanopillars (a) with crack width of $2c$ = 0.8 nm, 1.4 nm, 2.0 nm and 2.6 nm, and (b) with specimen widths of 20.0 nm, 30.0 nm, 40.0 nm and 50.0 nm, respectively. For all the cases in (b), the width of initial crack ($2c$) is 2.6 nm. The blue shadow regions on the left indicate the original elastic dilation of crack.

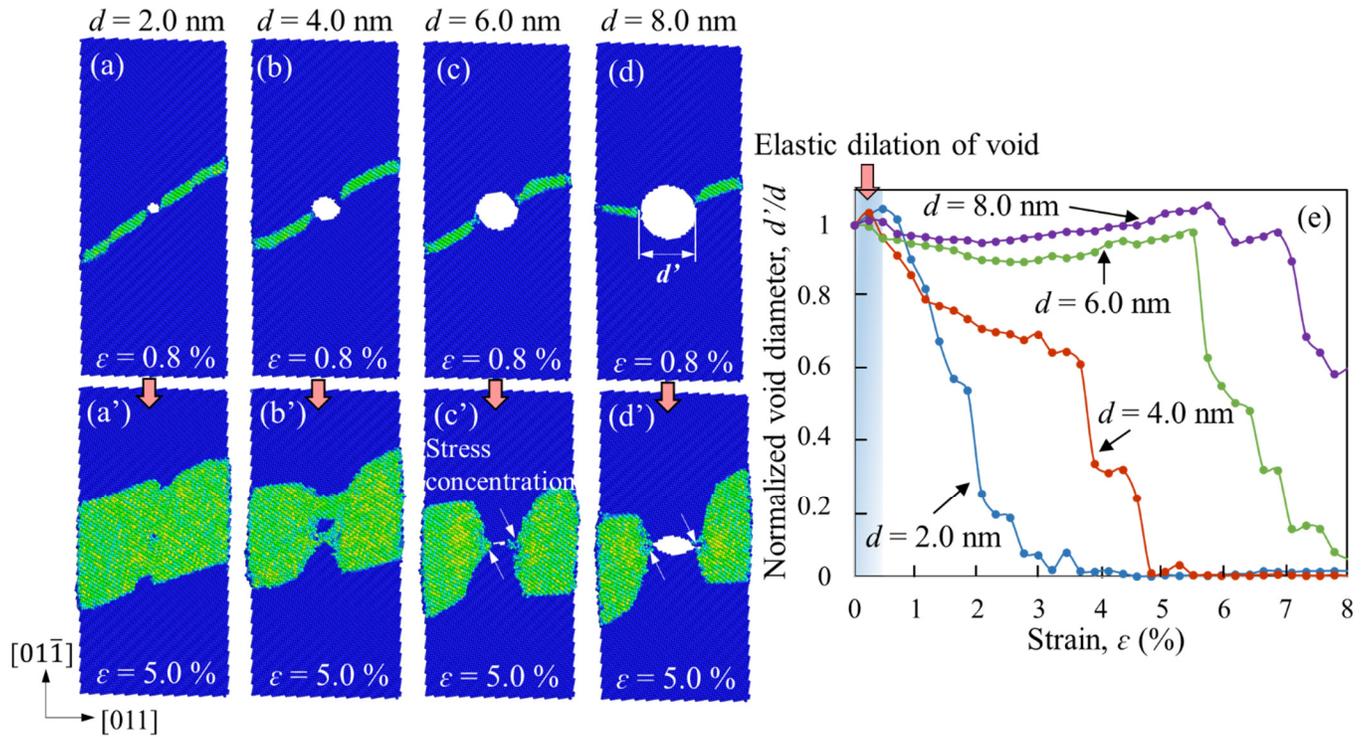

Figure 4. Atomic configurations of $[01\bar{1}]$-oriented YSTZ nanopillars with different pre-existing void sizes under compressive strains of (a-d) 0.8% and (a'-d') 5.0%. The atoms are colored by coordination number, where blue and green colors denote atoms with tetragonal and monoclinic crystal structures, respectively. (e) Evolution of normalized void diameter ($d'/d$) during compression as a function of strain for YSTZ nanopillars with pre-existing diameters of 2.0 nm, 4.0 nm, 6.0 nm and 8.0 nm. The blue shadow region on the left indicates the original elastic dilation of crack.

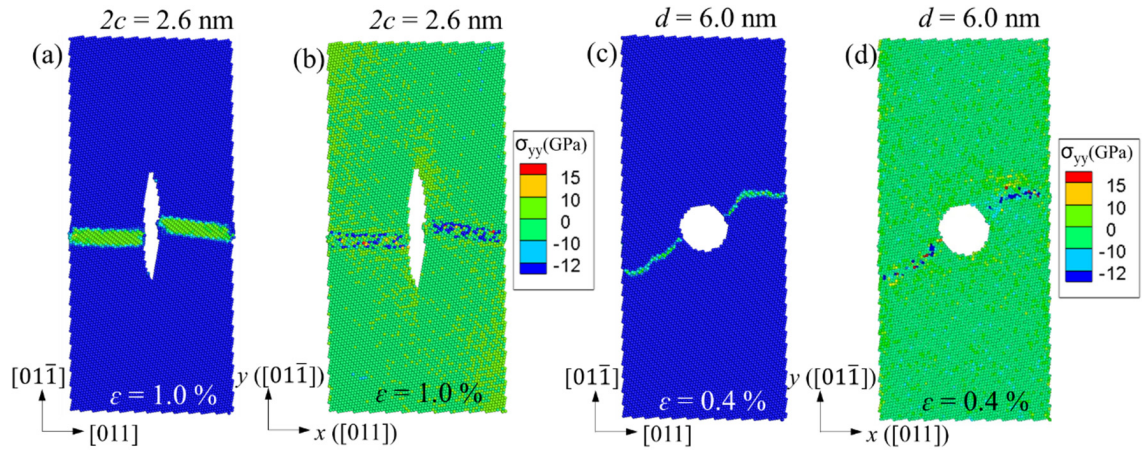

Figure 5. (a) Atomic configuration and (b) spatial distribution contour of $\sigma_{yy}$ of the $[01\bar{1}]$-oriented YSTZ nanopillars with pre-existing crack at strain of 1.0%. The crack has a width of 2.6 nm and a length of 18.2 nm. (c) Atomic configuration and (d) spatial distribution contour of $\sigma_{yy}$ of $[01\bar{1}]$-oriented YSTZ nanopillars with pre-existing void ($d = 6.0$ nm) at strain of 0.4%. Atoms in (a, c) and (b, d) are colored by CN and atom stress component of $\sigma_{yy}$, respectively.

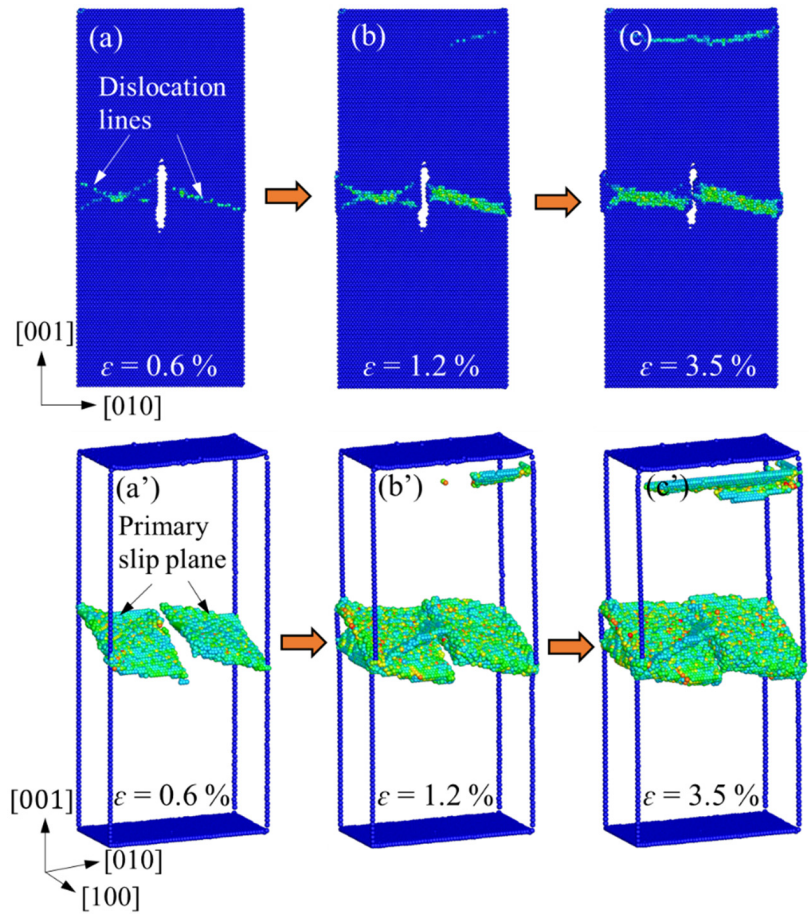

Figure 6. Atomic configurations of [001]-oriented YSTZ nanopillar with a pre-existing crack under compression. The length and width of crack are $2a$ = 9.8 nm, $2c$ = 1.4 nm. Atoms in (a-c) are colored by coordination number (CN). The slip planes in (a'-c') are revealed by slip vector (SV).